\documentclass[aps,prl,reprint,amsmath,amssymb,floatfix,nobibnotes]{revtex4-1}

\usepackage{graphicx}
\usepackage{dcolumn}
\usepackage{bm}
\usepackage[version=3]{mhchem} 
\usepackage{color}
\usepackage{textcomp}
\usepackage{float}

\begin{document}

\title{Enhancing terahertz generation from a two-color plasma using optical parametric amplifier waste light}

\author{Shayne A. Sorenson}\thanks{Authors contributed equally to this work.}
\author{Clayton D. Moss}\thanks{Authors contributed equally to this work.}
\author{Steven K. Kauwe}\thanks{Authors contributed equally to this work.}
\author{Jacob D. Bagley}
\author{Jeremy A. Johnson}\email{jjohnson@chem.byu.edu}
\affiliation{Department of Chemistry and Biochemistry, Brigham Young University, Provo, UT 84602}

\date{\today}

\begin{abstract}

We show experimentally that the terahertz (THz) emission of a plasma, generated in air by a two-color laser pulse (containing a near IR frequency and its second harmonic), can be enhanced by the addition of an 800-nm pulse.  We observed enhancements of the THz electric field by a factor of up to 30.  This provides a widely accessible means for researchers using optical parametric amplifiers (OPA) to increase their THz yields by simply adding the residual pump beam of the OPA to the plasma generating beam.  We investigate the dependence of the THz electric field enhancement factor on the powers of the two-color beam as well as the 800-nm enhancement beam.  Numerical calculations using the well-known photocurrent model are in excellent agreement with the experimental observations.   

\end{abstract}


\maketitle


The emerging applications of THz radiation in a wide array of technologies including chemical sensing and nondestructive material and biomedical imaging \cite{Dhillon2017,Yang2016}, as well as in fundamental materials research \cite{Hafez2016,Dastrup2017}, make the efficient generation of high-field THz radiation an ever-increasing need.  Especially desirable are strong table-top THz sources that would be widely accessible to researchers.  In addition to the more traditional techniques of photoconductive switching \cite{Jepsen1996} and optical rectification in nonlinear crystals \cite{Vicario2015,Lee2016}, plasma generation is also arising as a promising route to intense THz generation.  In short, this method typically involves combining an ultrafast laser pulse with its second harmonic in a tight focus to generate a plasma from which THz radiation is emitted \cite{Cook2000,Clerici2013,Dey2017}.  The THz radiation emitted from a two-color plasma is particularly attractive for its continuous, broadband spectral characteristics spanning from 0.1 THz to as high as 200 THz, depending on pulse duration\cite{Blank2013}.  Recent work by Kim and coworkers has shown that this emission is explained by a transient photocurrent model, wherein the nonlinearity of THz generation is due to the exponential nature of photoionization \cite{Kim2007,Kim2009}.

Having established the two-color filament generation technique, many researchers are actively investigating ways to optimize or improve the THz output \cite{Bagley2017,Kuk2016,Dai2007,Blanchard2009,Martinez2015,Lu2017,Zhang2017}.  Of particular relevance to this work, several researchers have explored the idea of using additional pulses prior to the two-color pulse in order to introduce a preexisting plasma and alter the THz emission \cite{Fan2013,Das2013,Xie2007, Minami2011}.  In one of these reports, Minami et al. showed that a prepulse had the clear effect of suppressing the THz generation of the plasma generation pulse; concluding that depleted neutral populations diminished the efficiency of plasma THz generation \cite{Minami2011}.In another vein of investigation, THz yields were drastically \emph{improved} using longer pump wavelengths, so much that the losses incurred during conversion in an OPA to longer wavelengths are more than made up for in the THz generation\cite{Clerici2013,Nguyen2016}.  Optimal table-top plasma THz generation is therefore best performed using an OPA.  This letter discusses a practical and value-adding use for the wasted residual OPA pump beam to enhance the THz yield of a two-color plasma.  We report on measurements showing that the THz emission of a two-color plasma is significantly enhanced, by as much as 30 times, when an additional 800-nm pulse is added to the plasma.  These experimental observations are corroborated by extending the transient photocurrent model to include three colors.


Terahertz radiation was generated using the two-color filamentation technique, as described previously \cite{Cook2000,Kim2007}.  Briefly, the near-IR output of an OPA (centered at 1450 nm with a pulse duration of about 100 fs) is passed through a BBO frequency doubling crystal and focused to form a plasma in dry air ($<2\%$ relative humidity).  The use of 1450 nm light balances the advantages of longer wavelengths \cite{Clerici2013} with the tuning curve of our OPA.  The emission of the plasma is collected and passed through a 1-mm Teflon filter, which rejects visible and near-IR frequencies and passes THz frequencies.  The THz electric field is measured using electro-optic sampling in 100 $\mu$m GaP \cite{Planken2001,Wu1997}. 

Initially, the horizontally-polarized THz emission of the two-color plasma was optimized without the enhancement beam by adjusting the polarization of the two-color pulse and the BBO phase matching angle and angle of incidence (which effectively tunes the phase relationship between the fundamental and the second harmonic).  The optimal polarization of the 1450-nm pulse was 30{\textdegree} from horizontal. The residual 800-nm beam from the second stage of a commercial OPA was used as the enhancement pulse, and we note that pumping the OPA results in a depleted beam with a non-gaussian spatial profile.  This enhancement pulse was collinearly added to the two-color plasma and its alignment and polarization were optimized for maximum enhancement.  The measured enhancements, as well as the optimum polarization of the enhancement beam were very sensitive to variations in alignment.  The likely optimum polarization for the enhancement beam was, nevertheless, found to be horizontal.  The enhancement pulse was set to a series of delays, relative to the two-color pulse, and a full measurement of the generated THz waveform was taken at each delay, averaging each point over 150 laser shots.  This was done for several combinations of powers of the enhancement and two-color beams, as well as for two polarizations of the enhancement beam.  THz yields for each measurement were calculated by taking the square root of the integrated power spectrum of each time trace (note that this quantity has the same scaling as the electric field strength).  Considering the Teflon filter, the 100 $\mu$m GaP and the duration of our EO sampling probe pulse, our detection bandwidth is limited to $<6$ THz.



\begin{figure}
  \includegraphics[width=3.4in]{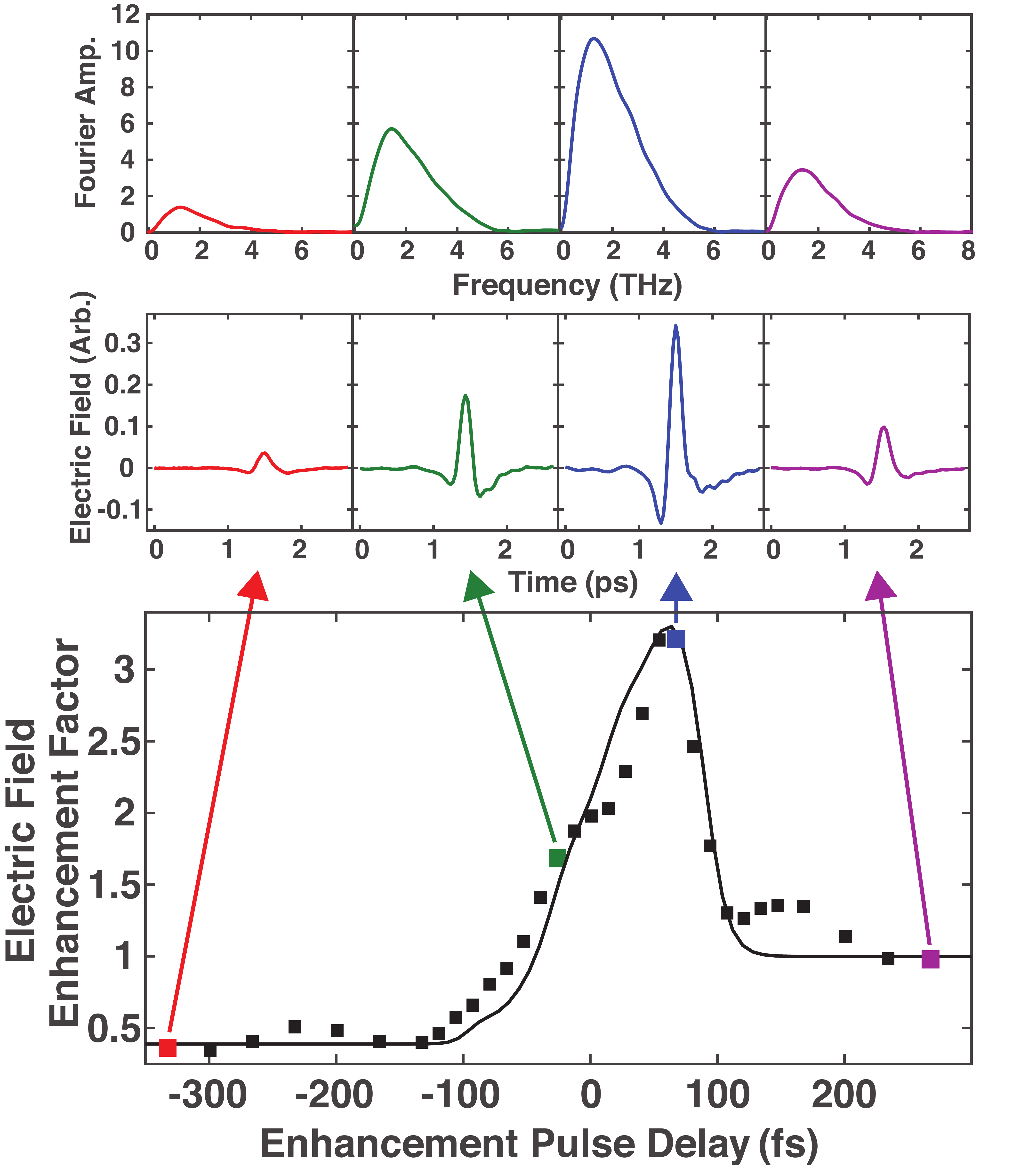}\\
  \caption{\label{figSingleEnhance} Bottom:  The measured (square symbols) and calculated (solid line) enhancement factor of the electric field strength of the plasma-generated THz as a function of delay of the 800-nm enhancement pulse relative to the two-color pulse.  The measured waveform, for the indicated points, along with their calculated Fourier amplitudes are shown above for the points indicated.}
\end{figure}

The critical finding of these measurements is that the 800-nm pulse, at a specific delay relative to the two-color pulse, \emph{enhances} the THz emission of the filament by as much as 30 times.  Our measurements also confirm that for negative delays (the enhancement pulse arrives before the two-color pulse) the THz yield is suppressed significantly, as demonstrated previously for the case where both the two-color pulse and the enhancement pulse were centered at 800 nm \cite{Fan2013,Minami2011,Das2013}.  Both of these findings are illustrated by plotting the THz yield of the filament as a function of the delay of the enhancement pulse relative to the two-color pulse.  For convenience, the THz yields can be considered relative to the yield in the absence of the enhancement pulse - we will call this the electric-field enhancement factor.  The bottom panel of Fig. \ref{figSingleEnhance} shows a representative plot of the electric-field enhancement factor as a function of enhancement pulse delay with two-color and enhancement pulse energies of 240 $\mu$J and 1930 $\mu$J, respectively.  Experimental measurements (represented by square symbols) are shown alongside model calculations (solid line) that are discussed below.  It is emphasized that each experimental point on these traces represents a full measurement of a THz waveform, i.e. for a given enhancement pulse delay:  the full THz waveform is measured, the Fourier transform is calculated and a THz yield is calculated as the square root of the integrated power spectrum.  Finally this THz yield is divided by the THz yield in the absence of the enhancement pulse to give a single point on the plot of electric field enhancement factors. In order to emphasize the actual measurements made, the top two rows of Fig. \ref{figSingleEnhance} show a selection of representative, measured THz waveforms and their Fourier transforms.  Importantly, the shapes of the spectra within our measurement window (0.1 - 6 THz) are consistent across the full range of powers we investigated, and no obvious spectral variations are introduced by the overlapping enhancement pulse\cite{SuppMat}.

\begin{figure}
  \includegraphics[width=3.4in]{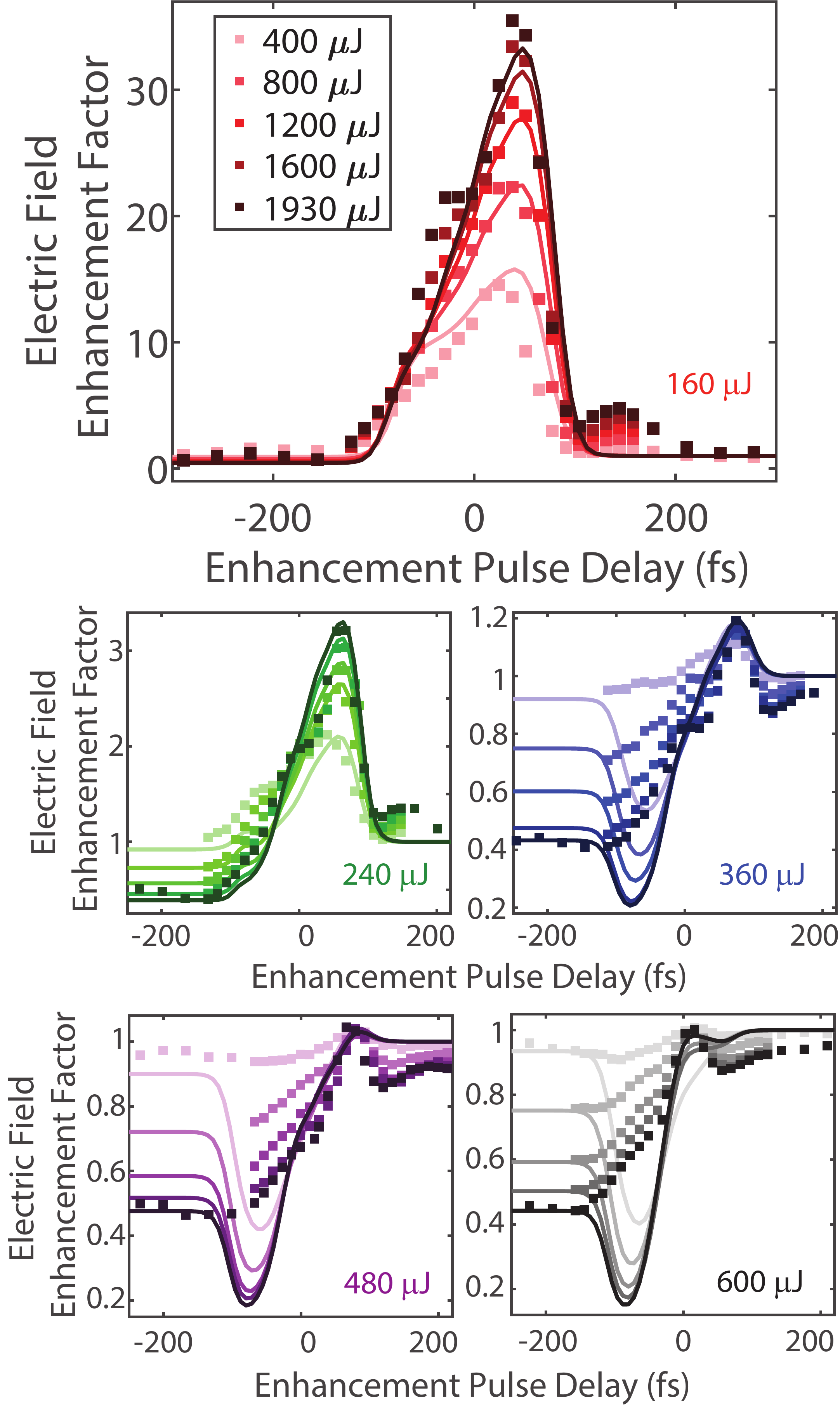}\\
  \caption{\label{figEnhanceWModel}Electric Field enhancement factors measured as a function of the delay of the enhancement pulse relative to the two-color pulse for several energies of the two-color pulse and the enhancement pulse.  Each pane contains measurements using a different two-color pulse energy as indicated, while the different traces represent different enhancement pulse energies.  The square symbols in each pane correspond to experimental measurements, while the solid lines correspond to numerical calculations using the plasma current model.  Of particular importance, are the enhancements achieved near a delay of 75 fs, reaching 30 times enhancement where the two-color pulse energy is 160 $\mu$J in the top pane.}
\end{figure}

Fig. \ref{figEnhanceWModel} shows plots of the electric field enhancement factor as a function of enhancement pulse delay for several two-color and enhancement pulse energies.  Each panel in Fig. \ref{figEnhanceWModel} shows a different two-color pulse energy, while the different shades within a single panel represent varying enhancement pulse energies.  Again, experimental measurements (represented by square symbols) are shown alongside model calculations (solid line).

Negative enhancement pulse delays in Fig \ref{figEnhanceWModel} correspond to the enhancement pulse arriving and generating the plasma \emph{before} the two-color pulse.  It is easily seen that at significant negative delays, the THz yield is substantially lower, or suppressed, relative to the yield of the two-color pulse alone.  On the other hand, at significant positive delays, the enhancement pulse arrives at the plasma \emph{after} the two-color pulse and the THz yields are equal to the yield of the two-color pulses alone.  For some combinations of input pulse energies, a significant increase of the THz yield is seen in Fig. \ref{figEnhanceWModel} at a small positive delay around 75 fs.

A few important observations can be made concerning the electric field enhancement measurements:  1)  At the lowest two-color pulse energy we studied, the THz emission is enhanced by as much as 30 times.  2)  This enhancement decreases with increasing two-color pulse energy, asymptotically approaching a value of 1 (corresponding to no enhancement) at two-color pulse energies around $600$ $\mu$J.  3)  For any given two-color pulse energy, the enhancement appears to saturate at high enhancement pulse energies.  Furthermore, the onset of this saturation appears earlier for the higher two-color pulse energies.  It even appears that at the highest two-color pulse energies (purple and black traces) saturation has already been reached at the lowest enhancement pulse energy, and no significant increase in THz yield is observed.  4)  Suppression of the THz yield at large negative delays varies between 0.9 and 0.4, and appears to correlate well with the enhancement pulse energy.  Suppression, however, does not appear to depend on the two-color pulse energy, as evidenced by the fact that the enhancement factors are roughly equal at the most negative delays across each pane of Fig. \ref{figEnhanceWModel}.


In order to explain these observations, we turn to the transient photocurrent model, which is presented in detail in Refs. \onlinecite{Kim2007,Kim2009}, and has been met with considerable success in describing the emission of two-color plasmas \cite{Nguyen2016,Vvedenskii2014,Borodin2013}.  The essence of the model is to consider the photo-acceleration of electrons liberated by tunneling ionization.  Details regarding our implementation of this model and an extension to include three colors, can be found in the supplemental materials \cite{SuppMat}.


The results of our calculations, shown alongside the measurements in Fig. \ref{figEnhanceWModel}, agree well with earlier reports showing a long-lived suppression of THz emission when the two-color filament is delayed with respect to a prepulse.  These observations and calculations are consistent with the understanding that the depletion of neutral gas molecules, caused by an intense prepulse, leads to the suppression of THz emission at negative delays\cite{SuppMat,Fan2013,Minami2011}.  Considering the need for larger THz electric fields, we find it more interesting to note that the model also captures the maximum enhancement expected for a given combination of prepulse and two-color pulse energies.  This can be seen for each of the curves in Fig. \ref{figEnhanceWModel} around a delay of 75 fs, where the maxima of the experiment and the model agree well.  The experiment and model also show that the dependence of the maximum enhancement on prepulse energy appears linear up to a point where it then saturates and no further increase in THz emission can be achieved.  The model suggests that this is due to saturation of the plasma itself since the number density of the gas is finite.  The physical underpinning for such large enhancements lies in the nonlinearity of the tunneling rate and the complexity of the interference produced by three incommensurate fields.  While both the two-color field and the 800-nm enhancement field are alone strong enough to ionize a significant number of carriers, when they are combined such that they constructively interfere, very large ionization jumps occur.  It is these steplike jumps that lead to the creation of new low frequency components in the emitted field \cite{Kim2007,Martinez2015}.  Interestingly, in the case of two-color plasmas, the relative phase of the fundamental and second-harmonic is critical for efficient conversion.  In contrast, for this case of incommensurate wavelengths the phase is not critical to enhancement provided the pulse durations are larger than the period of the beat frequency\cite{SuppMat}.   

In addition to capturing the suppression and enhancement of THz emission and their dependencies on both prepulse and two-color pulse energies, the model also captures some finer details observed experimentally.  For example the asymmetry in the plots corresponding to $160$ $\mu$J and $240$ $\mu$J in Fig. \ref{figEnhanceWModel} appears to agree quite well between model and experiment.  The experimental measurements show a small level of suppression or enhancement before the main features captured by the model - one likely explanation for this discrepancy is that there may be undesired pulses present in the experimental pulse train due to double reflections within the optical setup.  The excess suppression in the calculations at small negative delays also has a possible explanation.  Our method of performing an equally weighted average over a full cycle of relative enhancement pulse phases in the model calculations is roughly the equivalent of assuming a phase instability $>2\pi$ in our experimental apparatus on the time scale of {\raise.25ex\hbox{$\scriptstyle\sim$}}1 sec.  This is perhaps an overestimate of the instability and the model does show pronounced sensitivity to relative phase in that exact range of enhancement pulse delays \cite{SuppMat}.

\begin{figure}
  \includegraphics[width=3.4in]{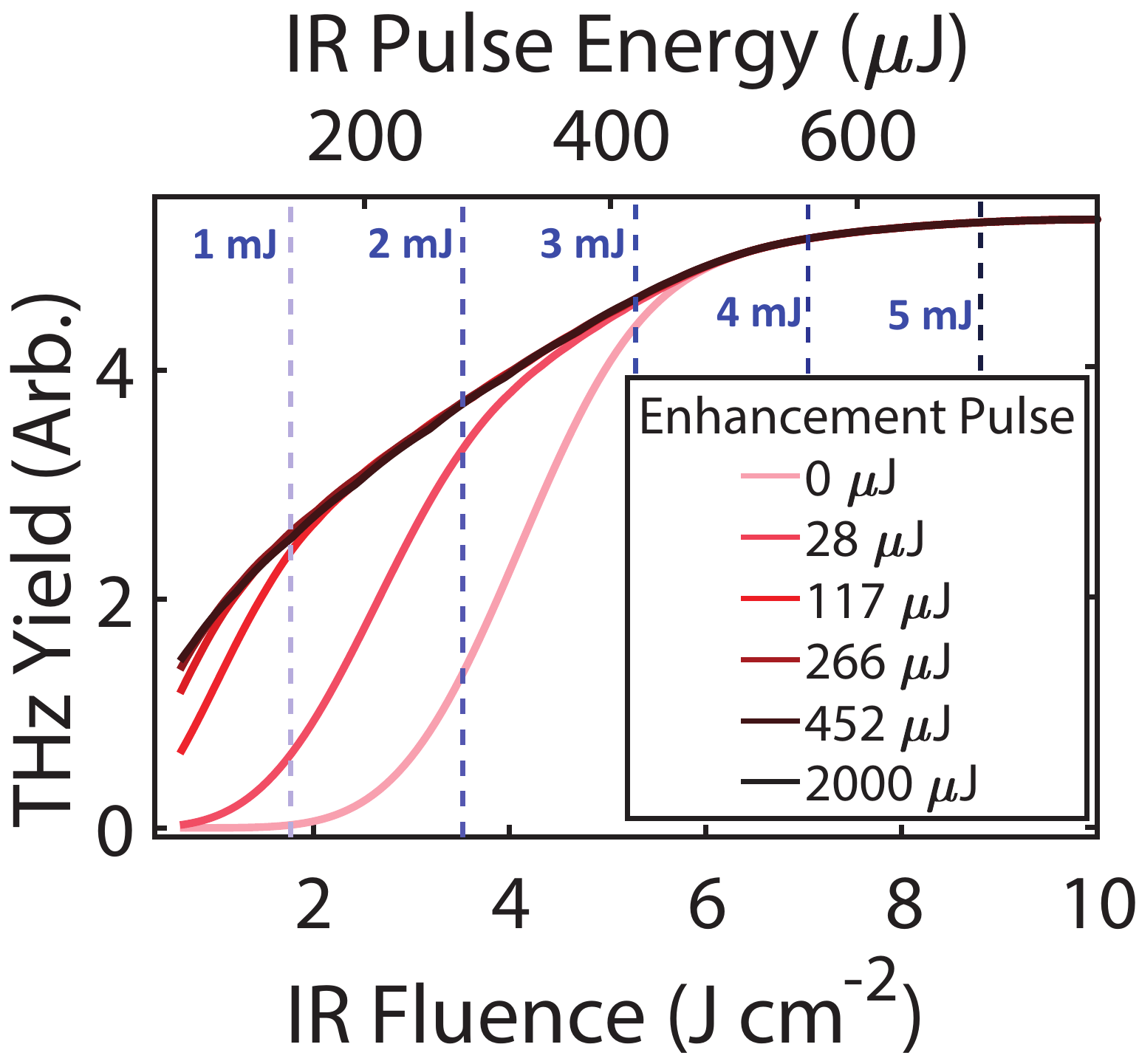}\\
  \caption{\label{figAbsYieldModel}Absolute THz yields (as opposed to relative enhancement factors) are shown as a function of enhancement pulse energy and two-color pulse energy.  The x-axis is two-color pulse energy and the different traces correspond to different enhancement pulse energies.  Large THz enhancements are seen at lower IR pulse energies, while saturation is apparent at higher IR pulse energies.}
\end{figure} 

Having established the veracity of the model calculations described, we can now use it to present a more complete and continuous picture of the observed THz enhancement.  Fig. \ref{figAbsYieldModel} shows the maximum expected THz yield, based on the model, as a function of two-color pulse fluence for several different enhancement pulse energies.  It is clearly seen how the addition of the 800-nm enhancement pulse can significantly enhance the THz emission of a filament, especially in cases where the fluence of the two-color pulse is relatively low.  A saturation of the THz output is observed where no further increases are obtained beyond about $600$ $\mu$J of IR pulse energy.  An important conclusion drawn from our results is that this saturation level depends on the number density of the neutral gas and the total optical fluence (the fluence of the two-color pulse plus that of the enhancement pulse).  In our current experimental configuration using a 5 mJ laser system, the saturation prevents us from achieving substantial improvements in the \emph{absolute} THz yield.  However, in many experimental implementations using different laser systems very large gains can immediately be made with a minimal amount of work.  Fig. \ref{figAbsYieldModel} includes lines, as guides to the eye, marking the expected enhancement for different laser system energies.  Bulk plasma effects such as the plasma frequency, opacity of the plasma and dispersion are not included in our model, nor have we scaled the absolute THz yield with the volume of plasma.  We anticipate that considering the complex interplay between optical fluence, plasma volume and the THz emission, there is likely a focussing geometry that would allow for a significant increase in \emph{absolute} THz yields even when using laser systems with $>5$ mJ pulse energies.  

We have demonstrated the ability to enhance THz emission from a two-color laser filament by as much as 30 times using traditionally wasted 800-nm light from an OPA.  Plasma generation of THz radiation has emerged as an important table-top THz source, and we anticipate that this simple means of enhancing that emission represents a fundamental advancement of the technique.  We emphasize that the enhancement of plasma emission is demonstrated here for the relevant case of a two-color plasma where the fundamental is a longer near-IR wavelength (1450-nm) in order to take advantage of the wavelength scaling of plasma generation\cite{Clerici2013}.  In light of recent work exploring the plasma generation of fractional harmonic frequencies \cite{Vvedenskii2014,Kostin2016,Zhang2017PRL}, in addition to our own results demonstrating THz generation using non-commensurate wavelengths\cite{Bagley2017}, it is especially interesting that this enhancement does not require a commensurate (or harmonic) frequency of the two-color pulse and is seemingly adaptable to a broad range of plasma generation frequency configurations.  A likely best practice for researchers desiring to generate large THz electric fields is to first convert to longer wavelengths using an OPA, in which case a large amount of optical power is typically wasted in the residual OPA pump beam.  We now understand, however, that this beam can be used in a very straightforward manner, with minimal cost, to further increase THz output.   

Importantly, the enhancement factor saturates at high optical fluences and oftentimes plasma generation of THz radiation is carried out in a regime of fluences where further enhancement can not be achieved.  Our results suggest that reducing the optical fluences of the two-color beam and the enhancement beam by increasing their spot sizes in the focal plane (essentially moving to the left in Fig \ref{figAbsYieldModel}) is a promising route to increased THz outputs.  Similar to reference \citenum{Kuk2016}, where THz emission is scaled up by increasing the plasma size, we hypothesize that the THz emitted from our three-color plasma could be significantly enhanced \emph{even at high IR pulse energies} by avoiding the carrier density saturation observed in the measurements reported here.  We also suspect that further gains can be made by increasing the density of the plasma supporting medium\cite{Dey2017}.  As mentioned above, we measured here the THz emission up to about 6 THz and were unable to comment on the emission outside of this range, however the full THz spectrum extending to 100 THz is also of great interest.  Further work will be directed to these ends.

\begin{acknowledgments}

This work was supported by the Department of Chemistry and Biochemistry and the College of Physical and Mathematical Sciences at Brigham Young University.

\end{acknowledgments}

\bibliography{Plasma800Enhancement}

\pagebreak
\onecolumngrid
\begin{center}
\textbf{Supplemental Material for:\\ Enhancing terahertz generation from a two-color plasma using optical parametric amplifier waste light}
\end{center}
\setcounter{equation}{0}
\setcounter{figure}{0}
\setcounter{table}{0}
\setcounter{page}{1}
\makeatletter
\renewcommand{\theequation}{S\arabic{equation}}
\renewcommand{\thefigure}{S\arabic{figure}}
\renewcommand{\bibnumfmt}[1]{[S#1]}
\renewcommand{\citenumfont}[1]{S#1}
\renewcommand{\thetable}{S\arabic{table}}%
\renewcommand{\thefigure}{S\arabic{figure}}%

\section{\label{secWaveforms}Experimental Setup}

\begin{figure} [H]
\centering
  \includegraphics[width=3.4in]{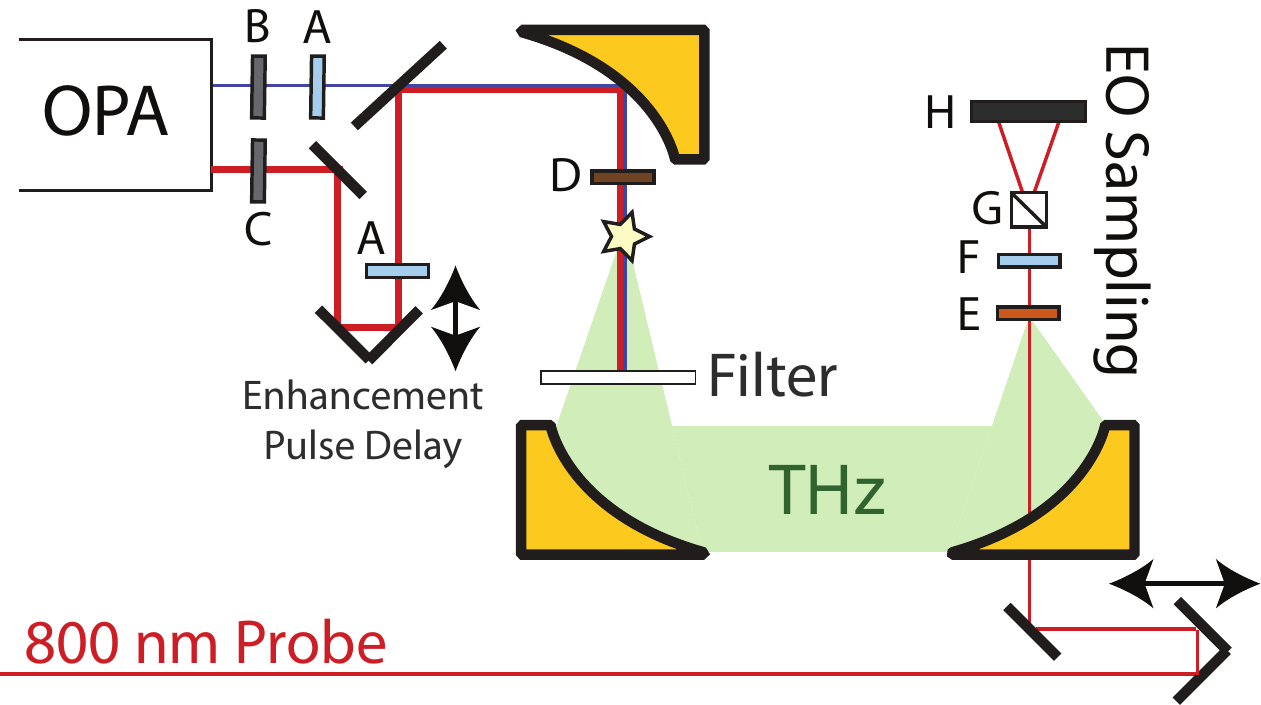}\\
  \caption{\label{figExpSetup}A schematic of the experimental apparatus.  The near IR output of an OPA is combined with the residual 800 nm pump of the OPA and focused in dry air to generate a plasma.  The THz emission of the plasma is collected and its electric field is measured using EO sampling on GaP.  The following optics are indicated:  A. $\lambda/2$ waveplate, B.  variable ND filter, C. polarizer pair, D. BBO crystal, E. GaP crystal, F. $\lambda/4$ waveplate, G. Polarizing beamsplitter, H. Balanced photodiode.}
\end{figure}

\section{\label{secMaxEnhance}Energy Dependence of Maximum Enhancement}

\begin{figure} [H]
\centering
  \includegraphics[width=3.4in]{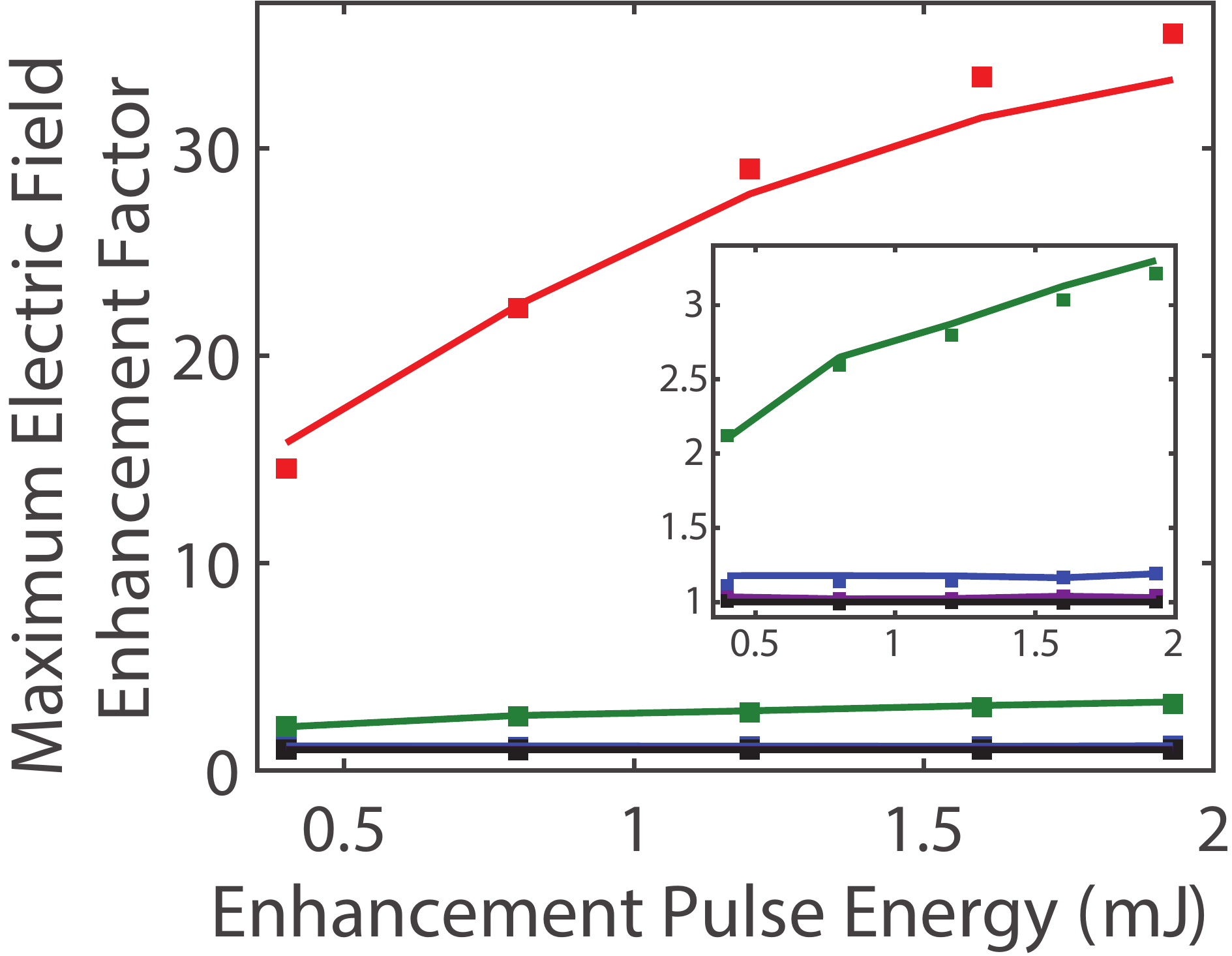}\\
  \caption{\label{figMaxEnhance}The maximum electric field enhancement from each trace in Fig. 2 of the main text is plotted as a function of enhancement pulse energy.  The different colors correspond to different two-color pulse energies and match the color scheme indicated in Fig. 2 of the main text.  Experimental measurements are indicated with square symbols and numerical calculations are indicated with solid lines.  The inset shows the same plot rescaled along the enhancement factor axis.  The dependence appears linear at lower enhancement pulse energies and then appears to start saturating at higher pulse energies.}
\end{figure}

Fig. \ref{figMaxEnhance} shows the maximum enhancement factor measured for each of the two-color and enhancement pulse energies.  Plotted this way, the saturation at higher total fluences may be easier to visualize.

\section{\label{secAlignment}Alignment Sensitivity}

\begin{figure} [H]
\centering
  \includegraphics[width=3.5in]{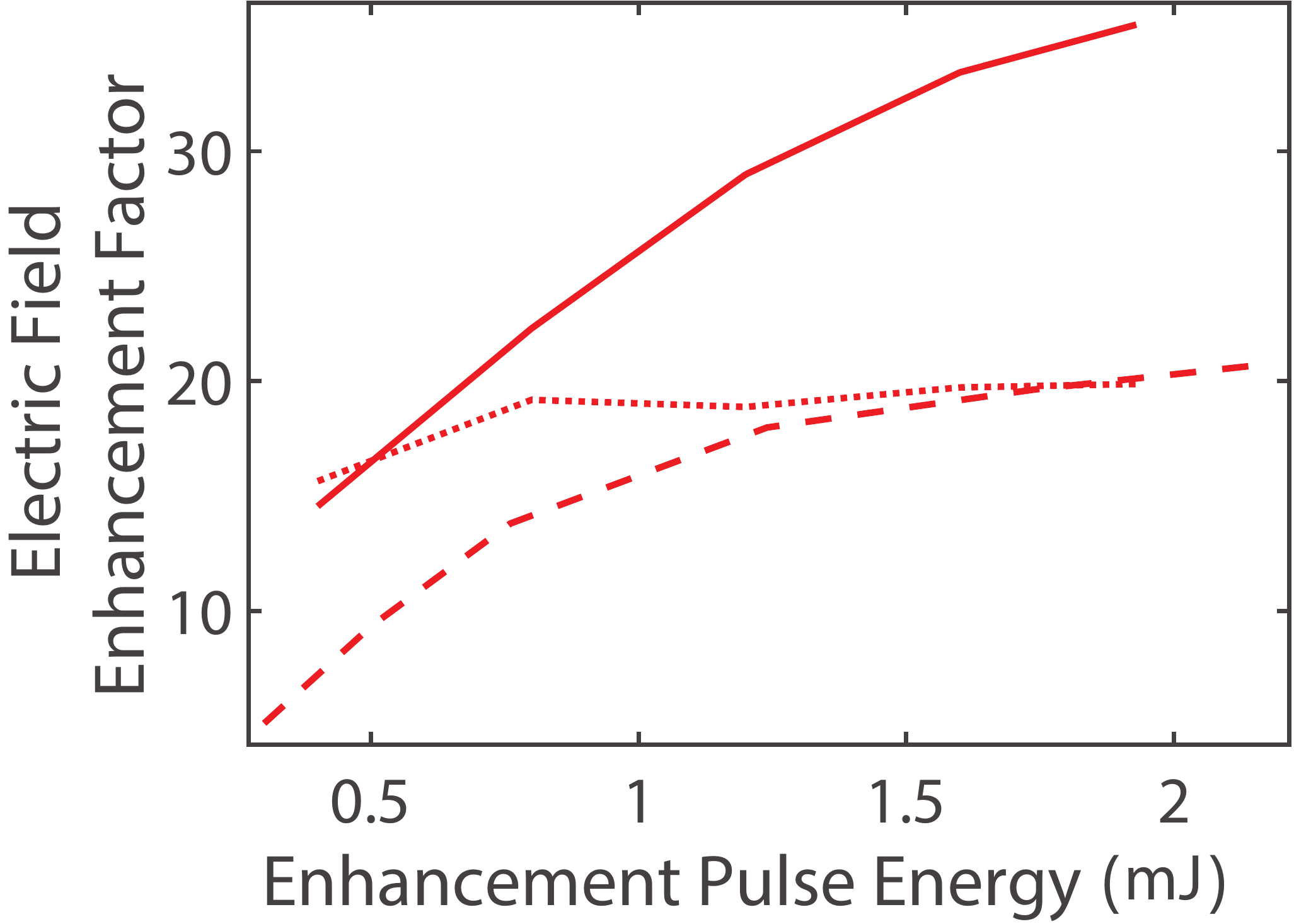}\\
  \caption{\label{figAlignment} Measured electric field enhancement factors for nominally the same two-color pulse energy of 160 $\mu$J on different days and presumably with slighly different beam alignment.}
\end{figure}

We observed that the electric field enhancements were dependent on alignment.  Despite best efforts to optimally align the enhancement beam for different sets of measurements, there remained some variability in the measured enhancment factors.  Along with this variability in acheivable enhancement, was a concomitant variability in the optimal polarization of the enhancement beam.  Fig. \ref{figAlignment} shows the maximum electric field enhancement factor as a function of enhancement pulse energy for three different sets of measurements taken on different days for nominally the same two-color pulse energy of 160 $\mu$J.  The solid red line is the same data shown in the main text and the measurements were taken with a horizontally polarized enhancement beam.  The two dashed lines were both taken with the enhancment beam polarized 25{\textdegree} relative to the horizontal axis.  In all three cases, the pointing and the polarization of the enhancement beam were adjusted to optimize the enhancement factor.  Furthermore, in all three cases the conclusions presented in the main text hold true concerning the magnitudes of the enhancements.

\section{\label{secResPol}Polarization dependence of the enhanced terahertz radiation}

\begin{figure}[H]
\centering
  \includegraphics[width=3.5in]{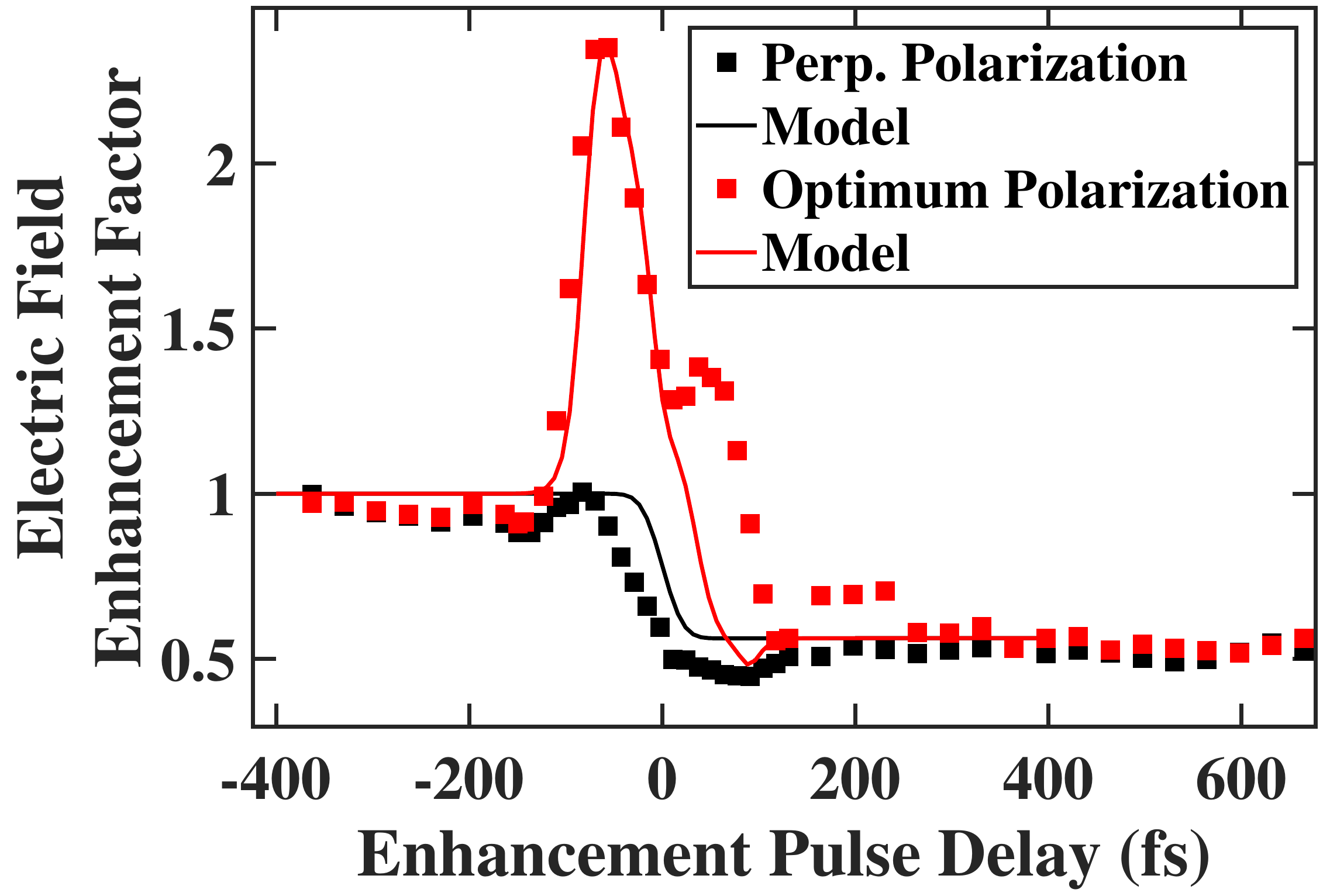}\\
  \caption{\label{figPolWModel}Electric field enhancement factors are shown for two perpendicular polarizations.  The measured (squares) and the calculated (lines) are in good agreement.  When the enhancement pulse is polarized perpendicular to the two-color field, no enhancement is observed.}
\end{figure}

As explained above, the measured enhancements, as well as the optimum polarization of the enhancement beam were very sensitive to variations in alignment, however the likely optimum polarization was found to be horizontal.  All the measurements shown in the main text were taken using the optimal enhancement beam polarization.  Electric field enhancement measurements were additionally taken with the enhancement beam polarization oriented perpendicular to the optimal polarization for a select number of the pulse energy combinations.  A representative comparison of the electric field enhancement for the optimal and the perpendicular polarization of the enhancement pulse is shown in Fig. \ref{figPolWModel}.  Again, the symbols represent experimental data and the solid lines show model calculations. 

We hypothesized that the THz yield for the measurements made with the perpendicular enhancement pulse polarization could be simulated by allowing the prepulse to assist in generating carriers but not accelerate them.  In other words, the field driving the tunneling ionization in the perpendicular polarization simulations is identical to that used in the main text including the fundamental, its second harmonic and the 800-nm pulses, but the field used in accelerating the electrons omits the 800-nm pulse contribution.  The results of such a calculation are shown in Fig. \ref{figPolWModel} along with the experiementally measured electric field enhancement factors. The important observations here are that the enhancement is practically eliminated when the prepulse polarization is rotated to be perpendicular to the experimentally determined optimal polarization.  It can also be seen that the suppression of the THz yield at larger negative delays appear to be identical between the two polarizations.

\section{Details on Modeling}

In our calculations, we have assumed that the number of free electrons satisfies the equation

\begin{equation}\label{eqdNdt}
\frac{dN(t)}{dt} = w(t)\left[ N_g - N(t) \right],
\end{equation}

where $N(t)$ is the total number of electrons at time $t$, $N_g$ is the number density of the neutral gas and $w(t)$ is the tunneling rate.  We have neglected recombination processes since these occur on timescales much longer than the pulse duration of both the pumping fields and the emitted THz field.  We used the static tunneling rate

\begin{equation}
w(t) = \frac{\alpha}{\hat{E}_L(t) } \exp \left( -\frac{\beta}{ \hat{E}_L(t) } \right),
\end{equation}

where $\alpha = 4\omega_a r_H^{5/2}$, $\beta = (2/3)r_H^{3/2}$, $\omega_a = 4.134\times10^{16}$ s$^{-1}$  is the atomic frequency unit and $r_H = U_{N_2}/U_{H}$ is the ionization potential of nitrogen gas (15.6 eV) relative to that of atomic hydrogen (13.6 eV).  We also define $\hat{E}_L(t) = \left| E(t) \right|/E_a$ as the absolute value of the electric field of the laser in atomic units ($E_a = 5.14\times10^{11}$ V/m).  We treat the electric field $E(t)$ as a linear combination of Gaussian pulses whose carrier frequencies and pulse widths are chosen to match the experimental conditions.  We define the phases and delays relative to the 1450 nm beam, and we assume the phase of the second harmonic, $\theta_{SH} = \pi/2$ in order to optimize THz yields in the simulations \cite{Kim2007}.  Tuning the angle of incidence of the BBO in the experiment to optimize the THz output is necessary in order to bring $\theta_{SH}$ close to this optimal value.  The amplitude of the electric field envelope for each of the three pulses was calculated based on the experimentally measured powers $P$ using

\begin{equation}
E_{max} = \frac{4}{d}\sqrt{\frac{P\sqrt{\ln{2}}}{\nu \pi^{3/2} \tau c\epsilon_0}},
\end{equation}   

where $d$ is the focal spot diameter, $\nu = 1$ kHz is the repetition rate of our laser, and $\tau = 75$ fs is the pulse duration.  In light of reports showing variability in the shape of a two-color plasma with changing pulse energies \cite{Oh2013,Theberge2006} we varied the spot diameters in the calculations.  The delay $\Delta$ between the enhancement pulse and the two-color pulse pair is determined experimentally and the phase of the enhancement pulse $\theta_{enh}$ was averaged in the simulations considering the distinct lack of phase stability in the experiments (further discussion on the effects of $\theta_{enh}$ can be found below).
  
Assuming that there are no free electrons in the gas before the pulses arrive (i.e. $N(t_0) = 0$), then the solution to Eq. \ref{eqdNdt} can be calculated as

\begin{equation}
N(t) = \exp{\left[ - \int_{t_0}^{t} dt' w(t') \right]} \times N_g \int_{t_0}^{t} dt' w(t') \exp{\left[ \int_{t_0}^{t'} dt'' w(t'') \right]  }
\end{equation}

The photo-emission of the plasma $E_{emit}(t)$, was then calculated as \cite{Borodin2013}

\begin{equation}\label{emit}
E_{emit} \propto \frac{e^2}{m_e}E(t)N(t),
\end{equation}
 
where $e$ and $m_e$ are the electron charge and mass, respectively.  As was done for the experimental measurements, THz yields were calculated from the simulated electric field time traces as the square root of the integral (from 0.1 to 6 THz) of the absolute value squared of the Fourier transform. The results of these calculations are shown alongside the experimental measurements for all the power combinations in Fig. 2 of the main text.  In using Eq. \ref{emit}, we have neglected the plasma oscillation and other bulk plasma effects which has been previously justified \cite{Andreeva2016, Debayle2014,Babushkin2010} and is also further discussed below.

\section{Spot Size Variation}

The beam diameter (and correspondingly, the diameter of the plasma) was expected to be close to 100 $\mu$m for all of the beams, however it was left as an adjustable parameter in light of the observation that filament size is distinctly dependent on the energy density of the laser pulses \cite{Theberge2006,Oh2013}.  In order to approximate the diameter of the plasma we performed a number of THz yield calculations using the plasma current model discussed in the paper over a range of enhancement and two-color beam fluences.  Since our model does not account for bulk plasma characteristics, the THz spot size only enters the model in calculating the electric field magnitudes from the pulse energies.  We can therefore use fluences in the model and avoid the explicit inclusion of the spot size.  

The calculations showed that the THz yield at significant negative enhancement pulse delays (where there is pronounced suppression of the THz yield) was independent of the two-color fluence.  This is corroborated by the experimental observations that are summarized in Fig. \ref{figSuppress} that shows the measured dependence of the suppression of THz emission as a function of enhancement pulse delay for the various two-color pulse energies.  We could plot a calculated curve of maximum suppression (or in other words minimum enhancement) vs. enhancement beam fluence to serve as a calibration curve where we look up the experimentally observed maximum suppression and find the corresponding enhancement beam fluence to use in the simulation.  This procedure using the suppression was critical in allowing us to separate the two pulse energies in correlating the calculations to the measurements. The procedure is illustrated in Fig. \ref{figSuppressFit}.

\begin{figure}[H]
\centering
  \includegraphics[width=3.5in]{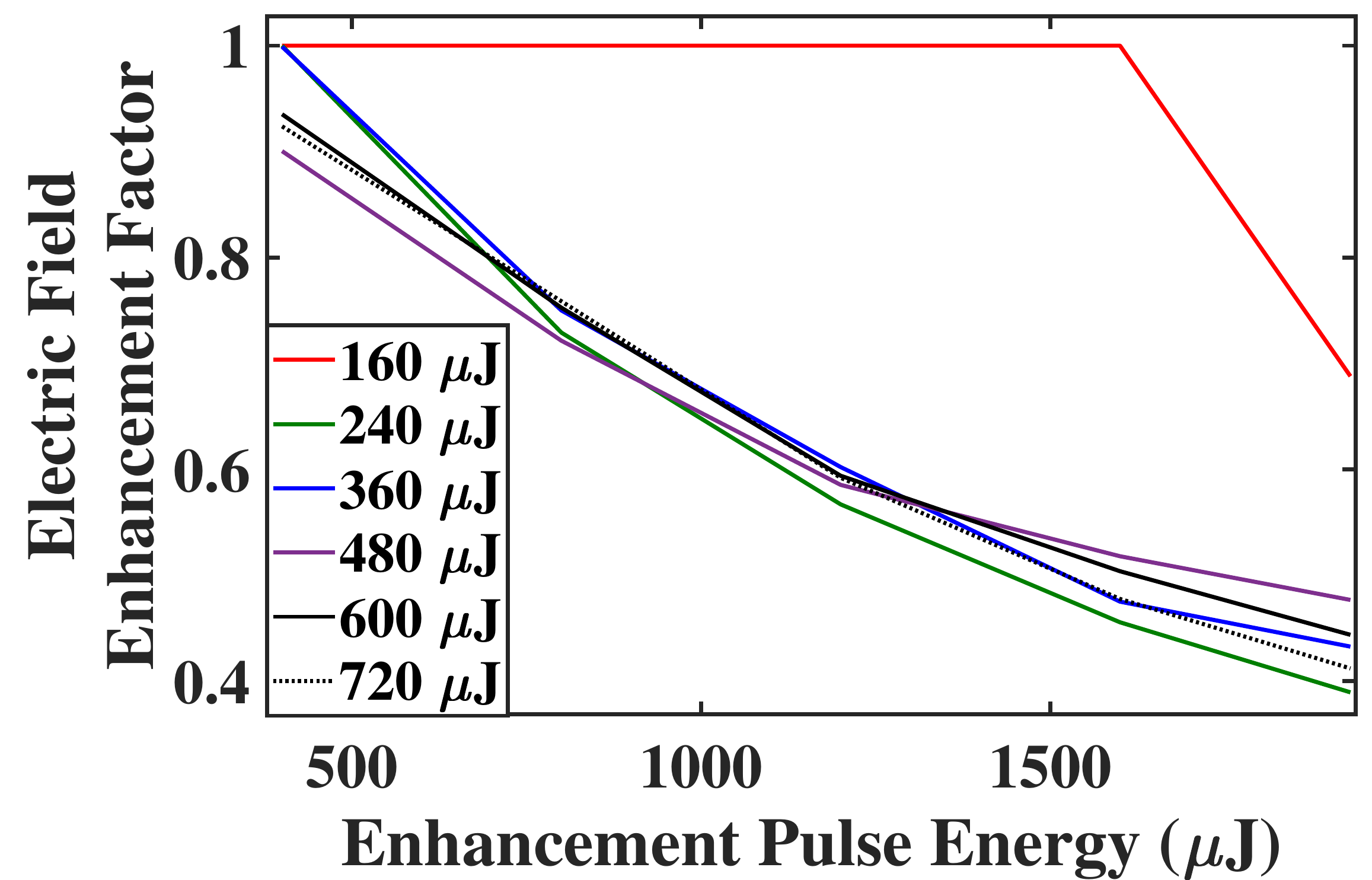}\\
  \caption{\label{figSuppress}The Experimentally measured electric field enhancement factor at a significant negative delay as a function of the enhancement pulse energy.  Of particular importance in the context of this section is that the shape and the magnitudes of the enhancement factors are essentially the same for all of the two-color pulse energies.  The 160 $\mu$J trace is an obvious outlier, because at lower enhancement pulse energies the measurements of the THz waveforms are approaching the noise floor of the electro-optic sampling apparatus and reliable measures of THz yield outside the region of enhancement (at a significant negative delay) were difficult to obtain.}
\end{figure}

\begin{figure} [H]
\centering
  \includegraphics[width=6in]{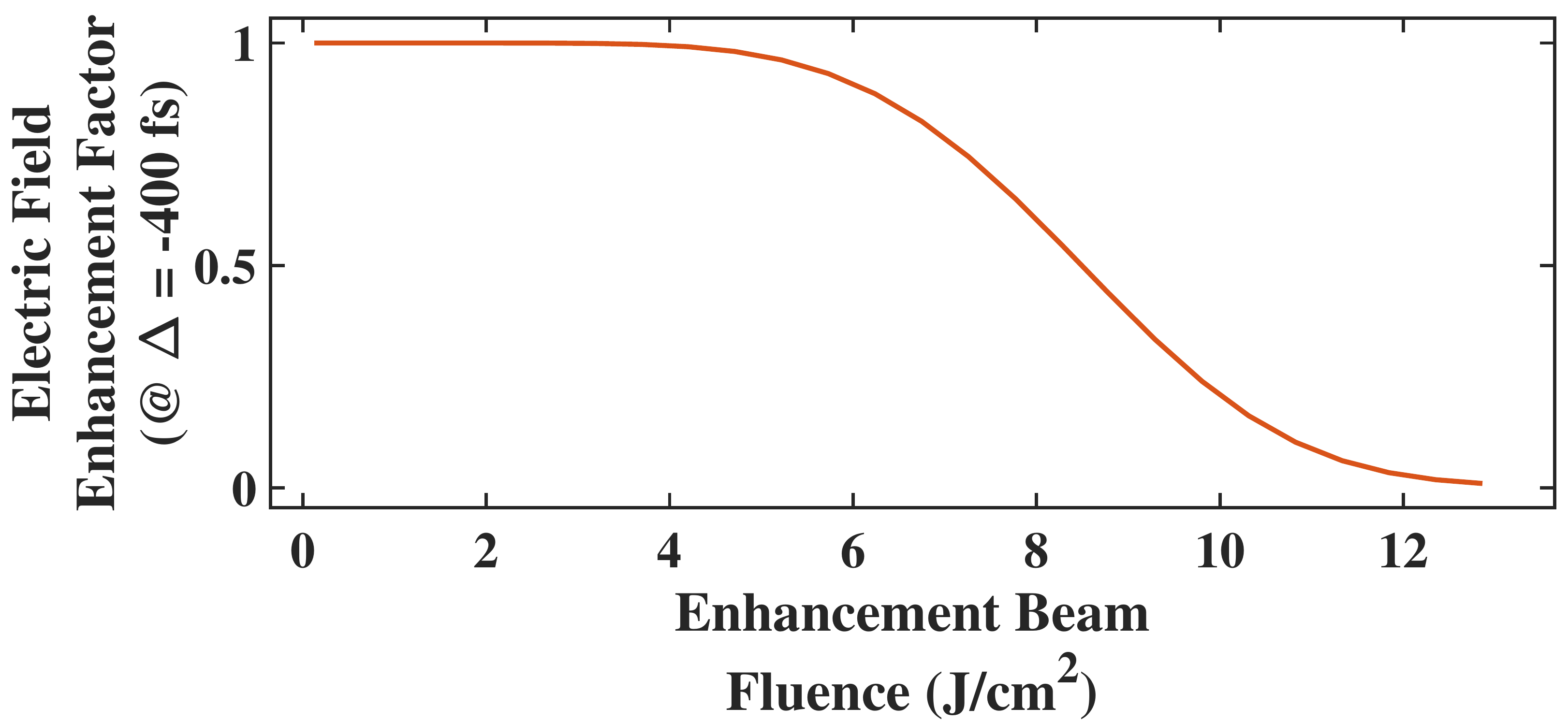}\\
  \caption{\label{figSuppressFit} Calculated enhancement factors at significant negative enhancement pulse delays were independent of the two-color fluence.  The curve shown here is true for all possible two-color fluences (within an experimentally conceivable range).  We were therefore able to match the measured suppression at significant negative delays and look those up on this curve to determine the enhancement beam fluence.}
\end{figure}

\begin{figure} [H]
\centering
  \includegraphics[width=3.5in]{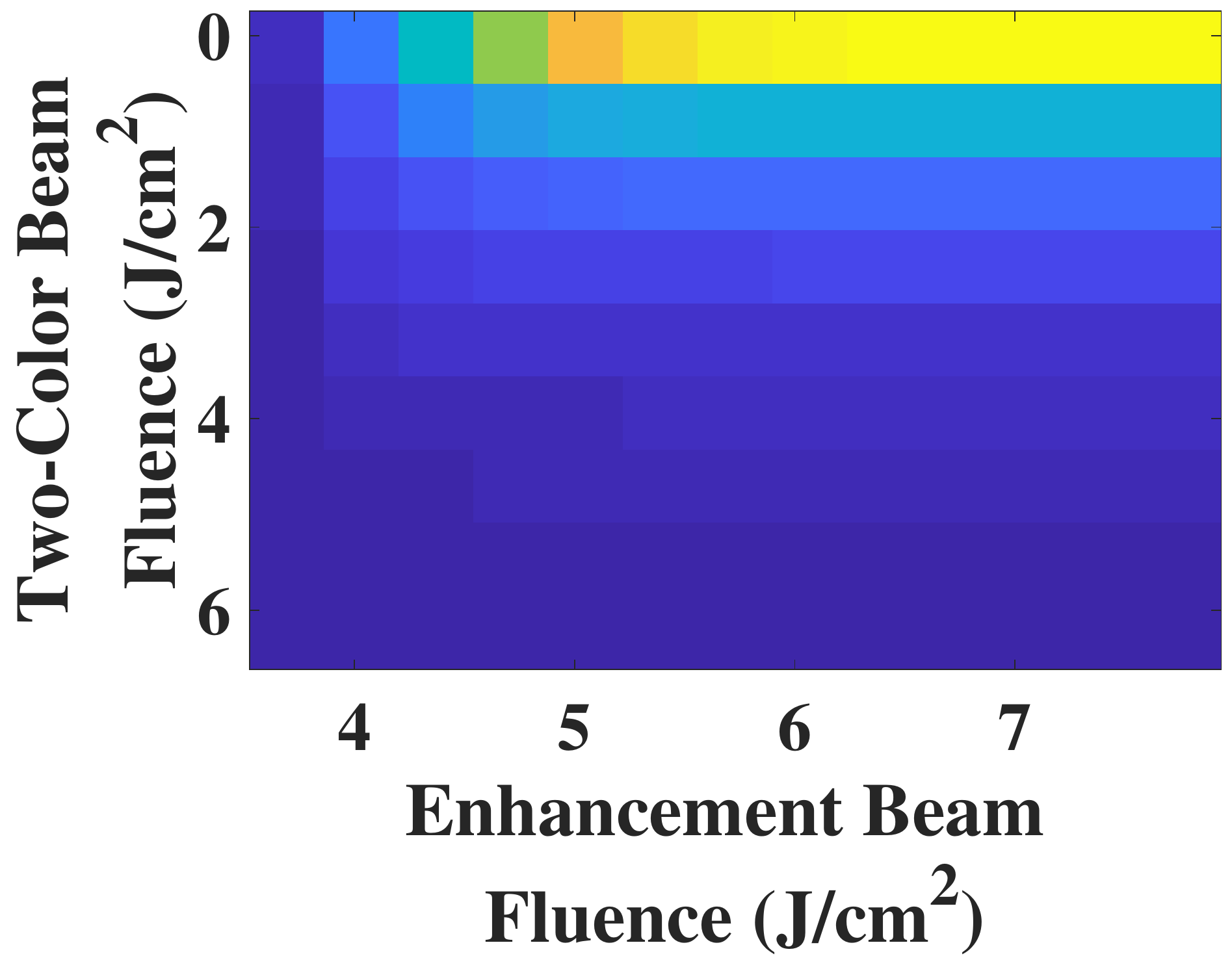}\\
  \caption{\label{figMaxFit} The maximum enhancement factor is a function of both the enhancement pulse and the two-color pulse energies.  Knowing the enhancement pulse energy from the suppression (obtained from Fig \ref{figSuppressFit}) allowed us to look on this surface, along the correct enhancement beam fluence, for the two-color beam fluence that would match the observed maximum enhancement.}
\end{figure}

Having determined the enhancement beam fluence, we were then able to determine a two-color beam fluence by looking up the experimentally observed maximum electric field enhancement factor on the calculated electric field enhancement factor surface.  The maximum enhancement factor is a function of both the enhancement and two-color beam fluences, thus the calculated maximum enhancement is a two-dimensional surface.  Once the enhancement beam fluence was determined from the suppression, the experimentally observed maximum enhancement could be matched to a single point on this surface and both the beam fluences independently determined.  Fig. \ref{figMaxFit} may assist in visualizing this process of determining the beam fluences using the maximum enhancement factor.

\begin{figure}[H]
\centering
  \includegraphics[width=6in]{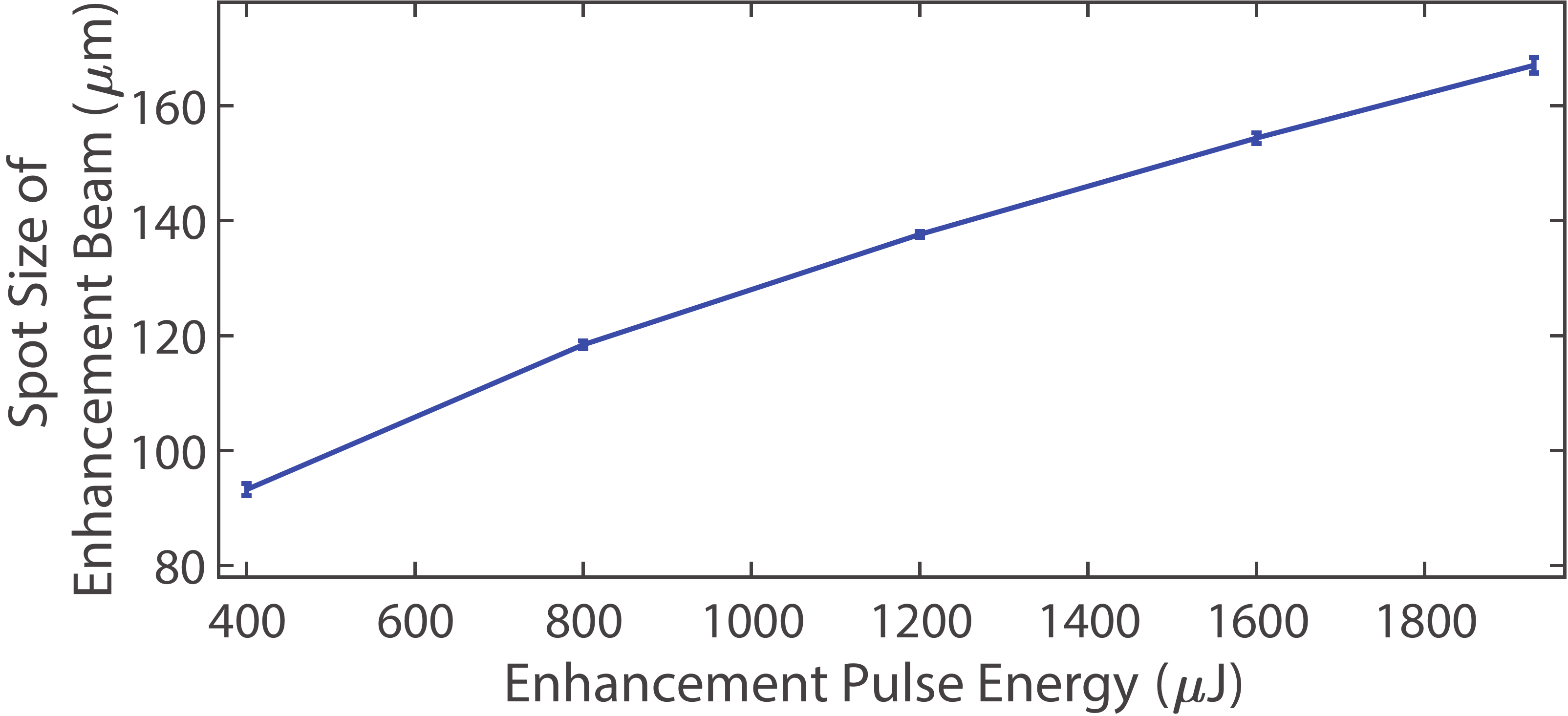}\\
  \caption{\label{figSpot800} The spot size of the enhancement beam as a function of the enhancement pulse energy determined by matching the calculations to the measurements.}
\end{figure}

\begin{figure}[H]
\centering
  \includegraphics[width=6in]{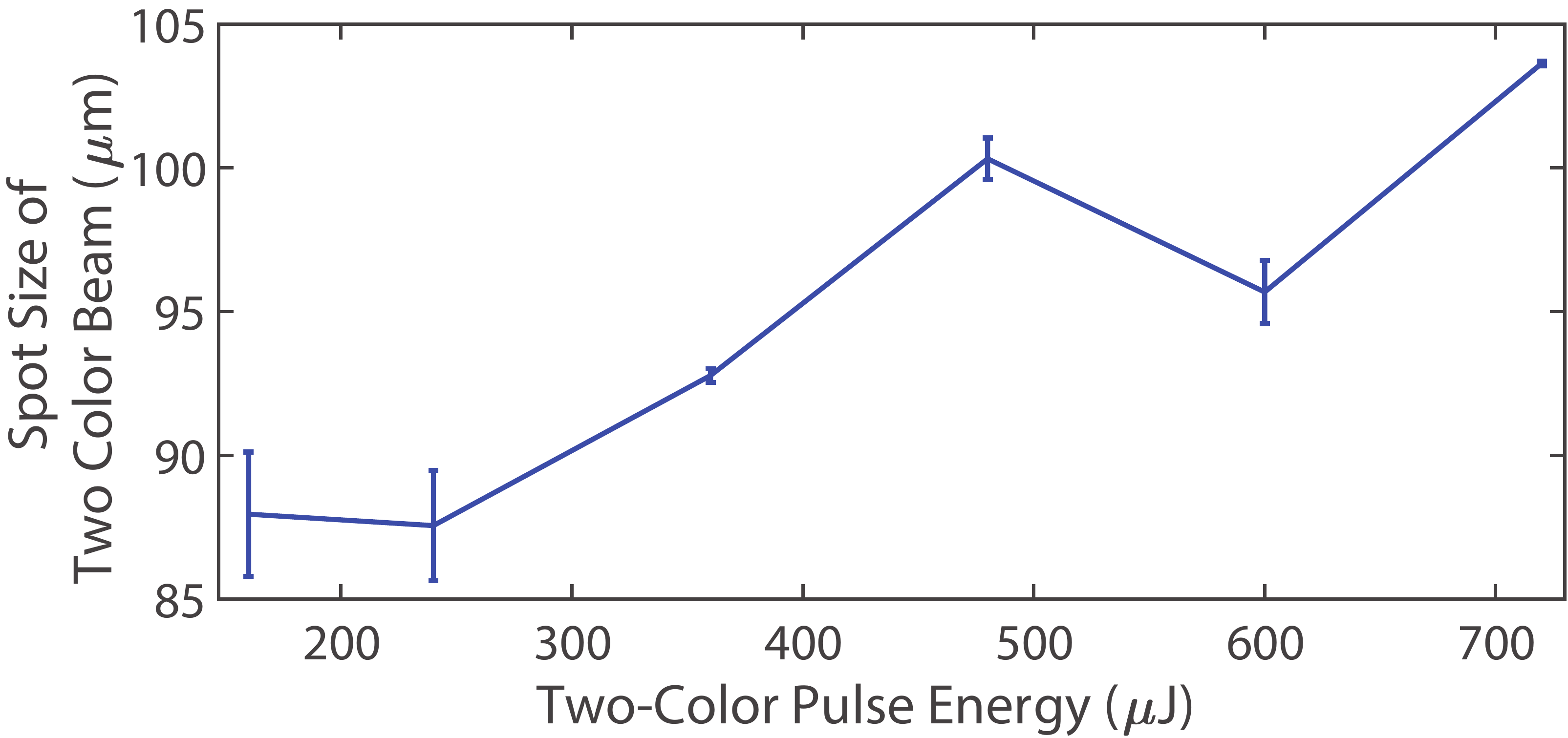}\\
  \caption{\label{figSpotIR}The spot size of the two-color beam as a function of the two-color pulse energy determined by matching the calculations to the measurements.}
\end{figure}

Finally we are able to determine spot sizes for both the enhancement and two-color beams since we have both a fluence (from matching calculation to experiment) and a measured pulse energy.  We find remarkable consistency of these calculated spot sizes accross the multiple measurements we performed.  The results presented in the letter include 5 measurements at each IR power and 6 measurements at each enhancement power.  Using these multiple measurements, we calculate a mean and standard deviation and plot the dependence of this theoretical spot size on pulse energy in Figs. \ref{figSpot800} and \ref{figSpotIR}.  It should be mentioned that these spot sizes and the general shape of the spot size curves is consistent with those measured directly in Ref .\cite{Theberge2006}.

\section{Phase Averaging}

\begin{figure}[H]
\centering
  \includegraphics[width=6in]{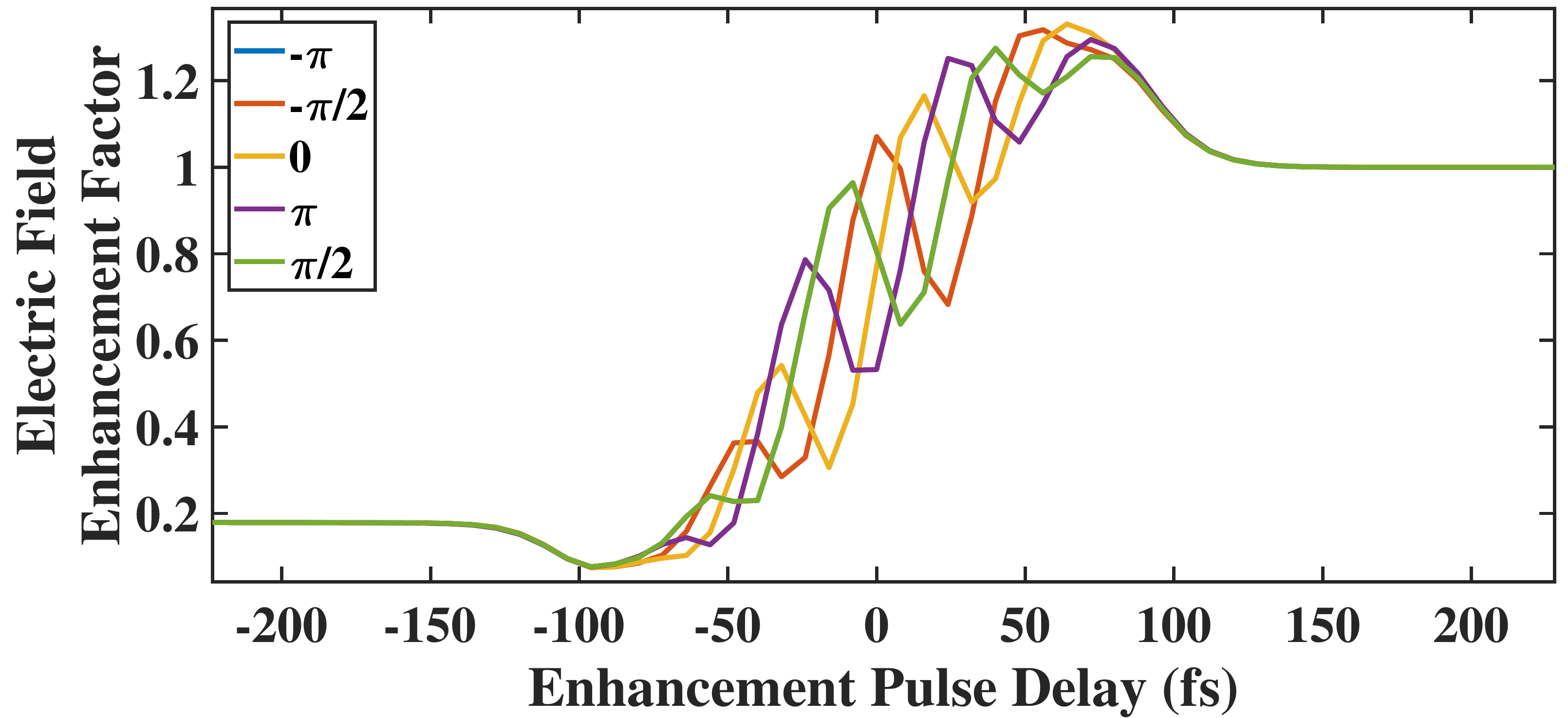}\\
  \caption{\label{figPhase800} The calculated electric field enhancement factor as a function of delay without averaging the enhancement pulse's phase.  Each line corresponds to the enhancement factor for a single enhancement pulse phase as indicated in the legend.  The phase of the second harmonic relative to the fundamental is still fixed at $\pi$/2.  The calculations in the main text used the average of an even distribution of several enhancement pulse phases from -$\pi$ to $\pi$}
\end{figure}

Experimentally, the fundamental pulse and it's second harmonic have a well-defined relative phase determined by the thickness of the BBO and the distance of the BBO from the plasma.  Since we use a white light seeded OPA pumped by a non-CEP stable Ti:Sapph laser, the carrier envelope phase of both the fundamental and its second harmonic vary from shot to shot.  Their relative phase however remains constant.  On the other hand, the 800 nm enhancement pulse and two-color beams follow different paths and consequently the prepulse phase $\theta_{pre}$ is not expected to be stable relative to the two-color pulse.  For this reason, in our modeling, the THz yield was calculated using an emission spectrum that was the equally weighted average of emission spectra calculated at 11 different phases between $-\pi$ and $\pi$.  This assumption of perfect phase instability may be a little too strong, which may explain the failure of the model calculations in exactly reproducing the measured enhancement factors at delays between 0 and ~100 fs.  Fig. \ref{figPhase800} shows calculated enhancement factor traces, again as a function of delay, but rather than averaging the phase of the enhancement pulse, we have just used a single phase.  It is seen how distinct oscillations appear in the plots in the region of delays between 0 and 100 fs and the phase of these oscillations depends on the chosen phase of the enhancement beam. 

\section{Number of Carriers}

Some physical insight into the mechanism of enhancement and suppression can be gained by considering the calculated traces of the number of carriers as a function of time.  It is already well understood that when the 800 nm enhancement pulse precedes the two-color pulse, suppression of the THz emission is due to the depletion of neutrals available to the two-color pulse.  This is illustrated in Fig. \ref{figSuppress} where the number of remaining neutrals in the simulations, immediately preceding the two-color pulse, is plotted as a function of the delay of the 800 nm enhancement pulse.  Negative delays correspond to the 800 nm pulse arriving before the two-color pulse and consequently the number of available neutrals is significantly reduced for negative enhancement pulse delays.  There is a clear, quantitative agreement between the enhancement pulse energy, the observed levels of suppression in Fig. 2 of the main text and the number of remaining neutrals at negative enhancement pulse delays, which illustrates the connection between the measured suppression and the depletion of neutrals by a prepulse.

\begin{figure}[H]
\centering
  \includegraphics[width=6in]{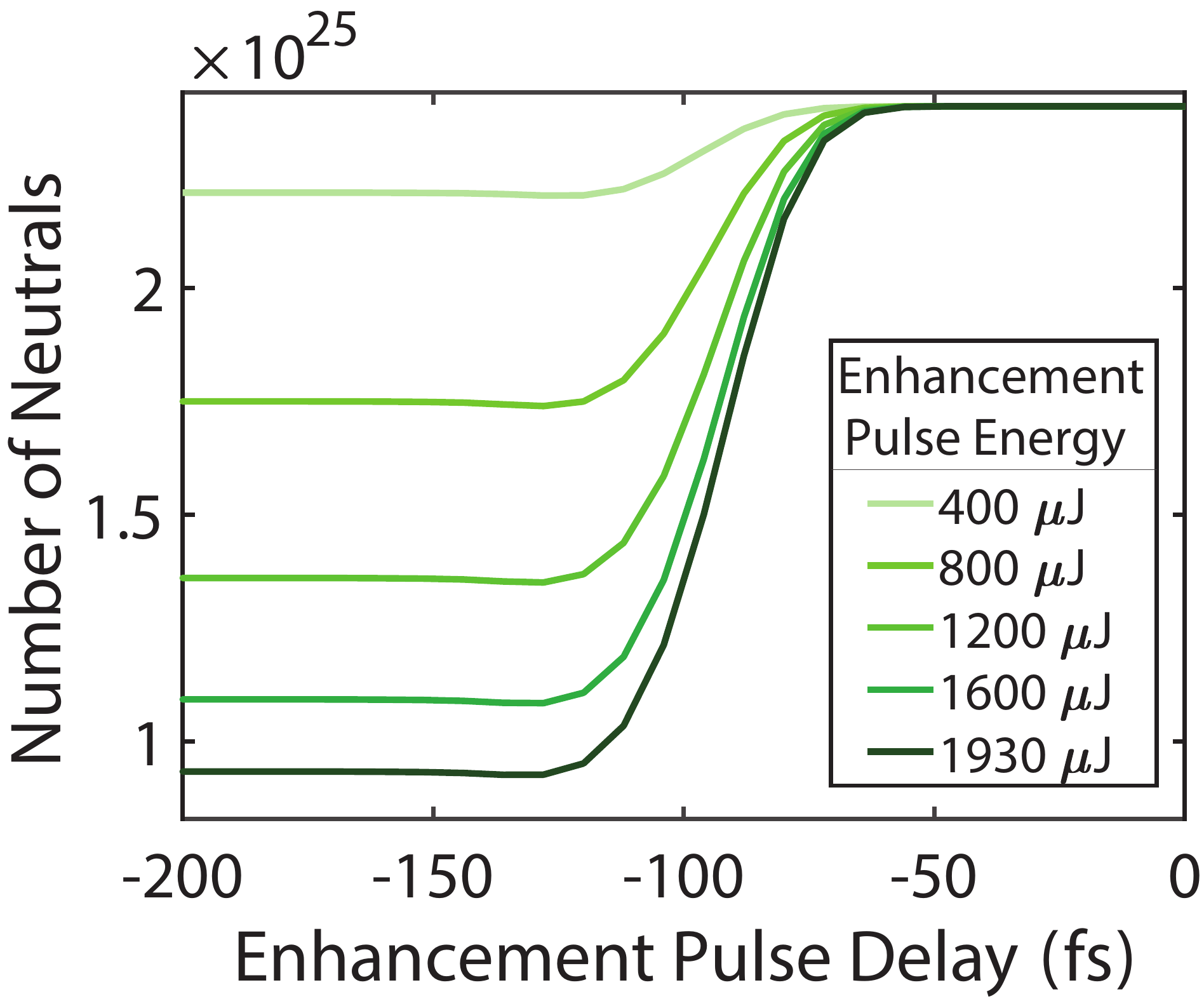}\\
  \caption{\label{figSuppress} The density of neutrals ($m^{-3}$), from simulations, immediately preceding the arrival of the two-color pulse.  Different traces correspond to different enhancement pulse energies and a clear correlation can be seen between the pulse energy, the observed levels of suppression in Fig. 2 of the manuscript and the number of neutrals here.}
\end{figure} 

\begin{figure}[H]
\centering
  \includegraphics[width=6in]{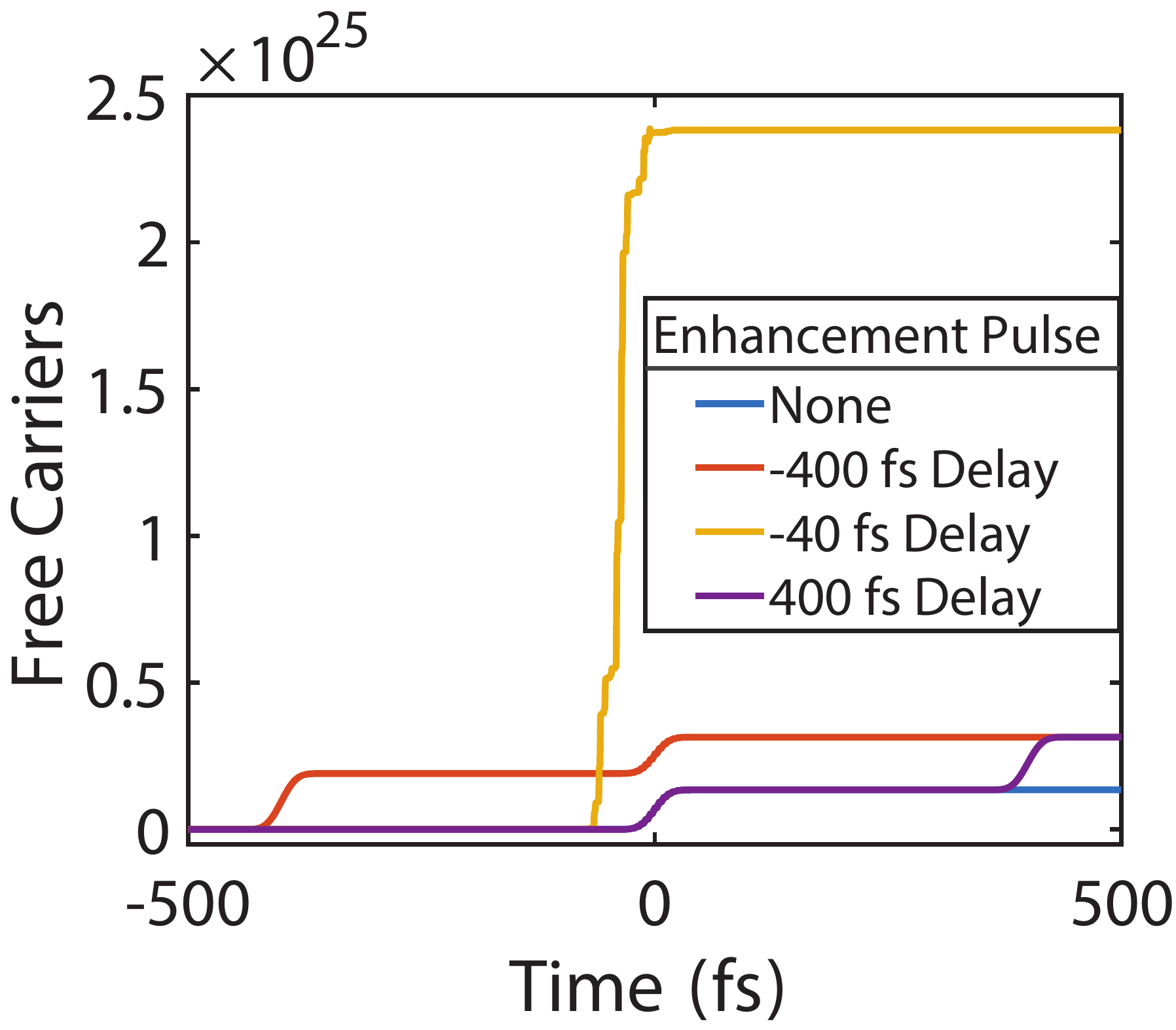}\\
  \caption{\label{figEnhance}Number of carriers ($m^{-3}$) from simulations as a function of time.  This clearly shows that when the enhancement pulse is overlapped with the two-color pulse, much larger step-like jumps in the number of carriers occur.  It is these jumps that lead to the creation of new frequencies in the emission spectrum.}
\end{figure}

In order to better visualize the mechanism of enhancement we consider the number of carriers in the plasma as a function of time.  Since the tunneling rate is highly nonlinear, the number of carriers produced by the addition of multiple driving fields depends sensitively on the phase and delay of those driving fields.  Fig. \ref{figEnhance} shows traces of the number of carriers in the plasma, from simulations, for a few different enhancement pulse delays, as well as in the absence of the enhancement pulse.  Here it is easily seen that when the two-color and enhancement pulses constructively interfere, the number of carriers generated makes much larger step-like jumps.  In these experiments, the enhancement pulse and the fundamental pulse are not harmonics of one another, so the addition of these two fields would result in a beat frequency.  Importantly, the period of the beat frequency is much smaller than the laser pulse durations so regardless of the initial phase between the two pulses, there will always be times of very large constructive interference and, consequently, very large, step-like jumps in the number of carriers generated.  This is the reason why phase stability of the enhancement pulse is not a strict requirement for enhancement.  Ultimately, as can be seen by taking the Fourier transform of Eq. \ref{emit}, it's these step-like jumps that lead to the production of new frequencies in the plasma emission. 

\begin{figure}[H]
\centering
  \includegraphics[width=5in]{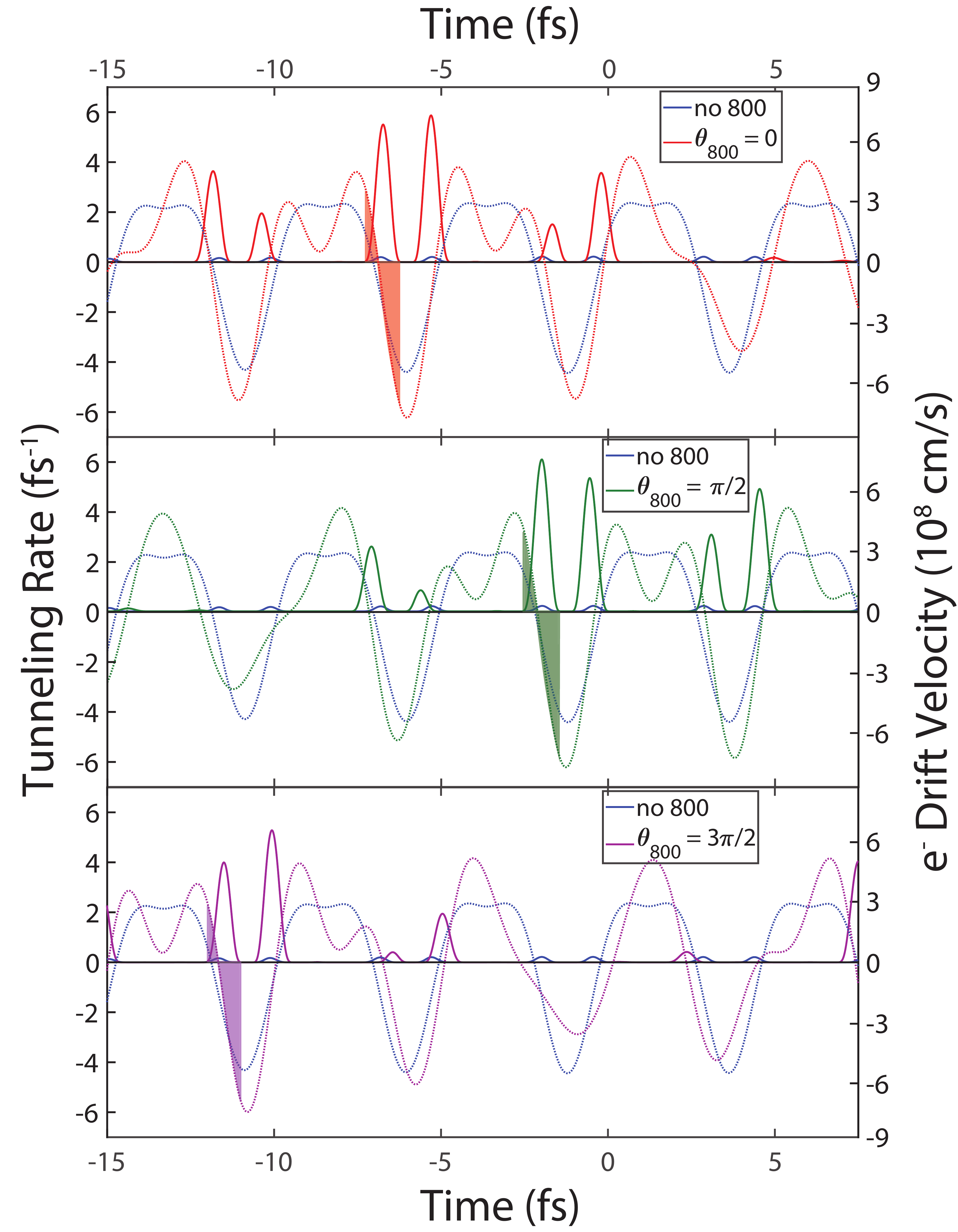}\\
  \caption{\label{figDemoPhysics}Each panel shows a plot of the drift velocity (dotted lines) of the carriers ionized by a three-color field, where the initial phase of the enhancement beam is indicated in the legends, as well as the tunneling rate (solid lines) for that same field.  For comparison, a trace of the drift velocity and the tunneling rate in the absence of the enhancement beam is also shown in each panel.  The asymmetry in the drift velocity during times of high tunneling rate are what lead to THz emission - this is highlighted for one such instance in each panel by the filled-in curves.}
\end{figure}

A similar argument to that made by Kim et al. in Ref. \citenum{Kim2007} concerning the carrier drift velocities can also be made in this case of 3-color plasmas.  The essence of the argument is that electrons liberated at any given time will have a directional drift velocity (in addition to the oscillatory velocity that follows the field).  As explained above, in our experiments the overwhelming majority of carriers are ionized at just a few times due to the high non-linearity of tunneling ionization.  Considering the drift velocities of carriers liberated at these times we see that for every phase of the 800 there is a distinct asymmetry and the majority of carriers are accelerated preferrentially in one direction.  This is illustrated in Fig. \ref{figDemoPhysics}, where the electron drift velocities (dotted lines) are shown as a function of the time at which they were liberated along with the tunneling rate (solid lines) as a function of time.  Each panel shows a different relative phase of the 800-nm enhancement beam.  The drift velocities and tunneling rates in the absence of the 800-nm enhancement beam are also shown in each for easy comparison.  It is readily seen that, regardless of the initial phase of the enhancement pulse, the coincidence in time of both very large step-like jumps in number of carriers and asymmetric drift velocity explain the THz emission of the three-color plasmas.  

\section{THz Spectra from Measurements}

\begin{figure}[H]
\centering
  \includegraphics[width=6in]{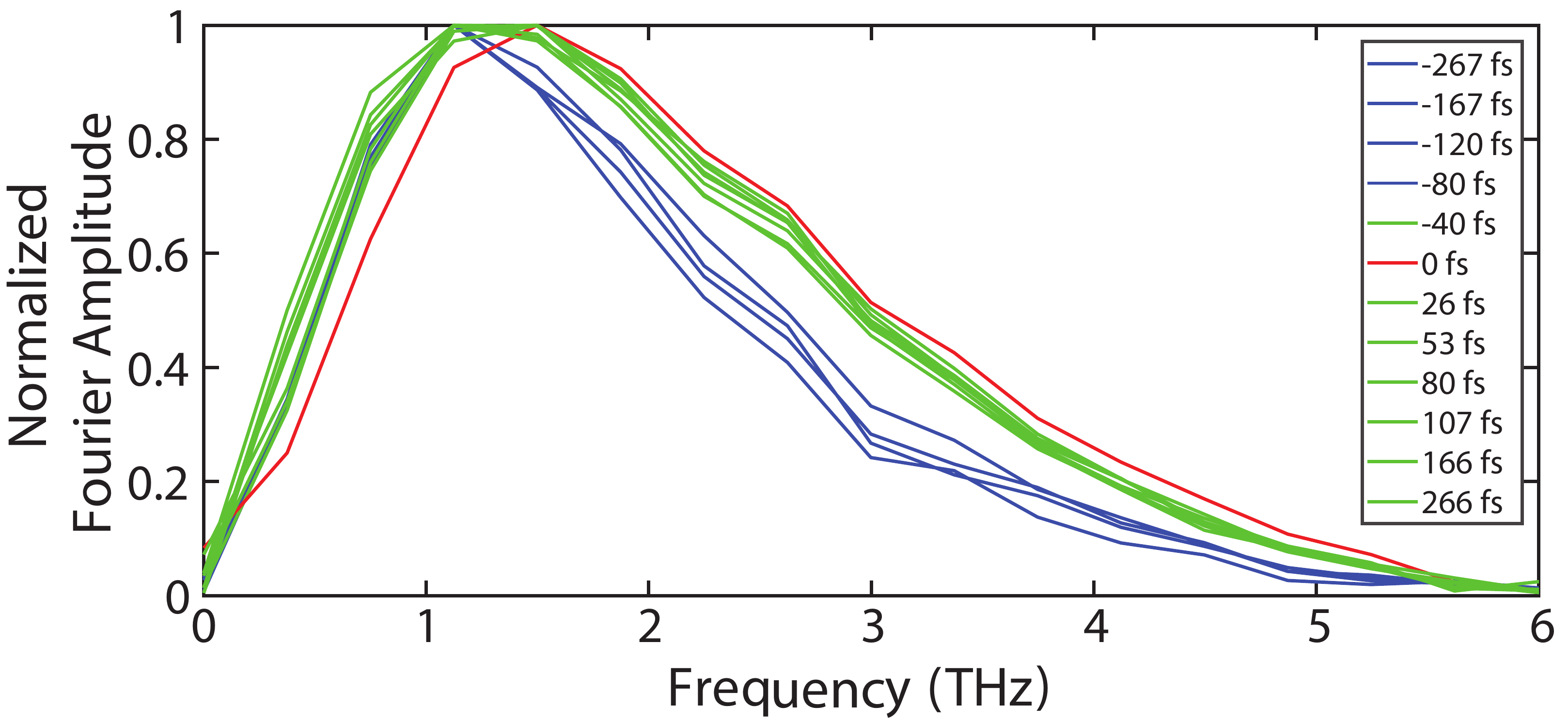}\\
  \caption{\label{figNormalizedSpectra} Measured, normalized spectra of the THz emission from the three-color plasma at several enhancement pulse delays.  The general shape of the spectra is consistent at all delays with small shifts or broadenings on the order of tenths of THz.  This consistency further justifies our neglect of the plasma frequency in our simulations.}
\end{figure}

In using Eq. \ref{emit}, we have entirely neglected the plasma frequency.  In addition to being consistent with previous work \cite{Debayle2014,Babushkin2010}, we can also justify this on two different grounds.  First, we look at the observed THz emission spectra and see that the shape is consistent accross the full range of powers and delays that we investigate.  In light of several works showing THz emission spectra peaking at the plasma frequency\cite{Andreeva2016}, we conclude that either the plasma frequency is not an important contribution in this regime of fluences and wavelengths, or that the plasma frequency is not changing much over our range of powers and delays.  Several representative normalized spectra of measured THz waveforms are shown in Fig. \ref{figNormalizedSpectra}, for the case of a two-color pulse energy of 240 $\mu$J.  Any shifts in the spectral shape are quite small relative to the window of integration used.  We do point out however that there seems to be a systematic shift on the order of a few tenths of THz in moving from delays where suppression is observed (negative) to delays where enhancement is observed (close to zero) to delays where the two-color pulse precedes the enhancement pulse (positive). 

\begin{figure}[H]
\centering
  \includegraphics[width=6in]{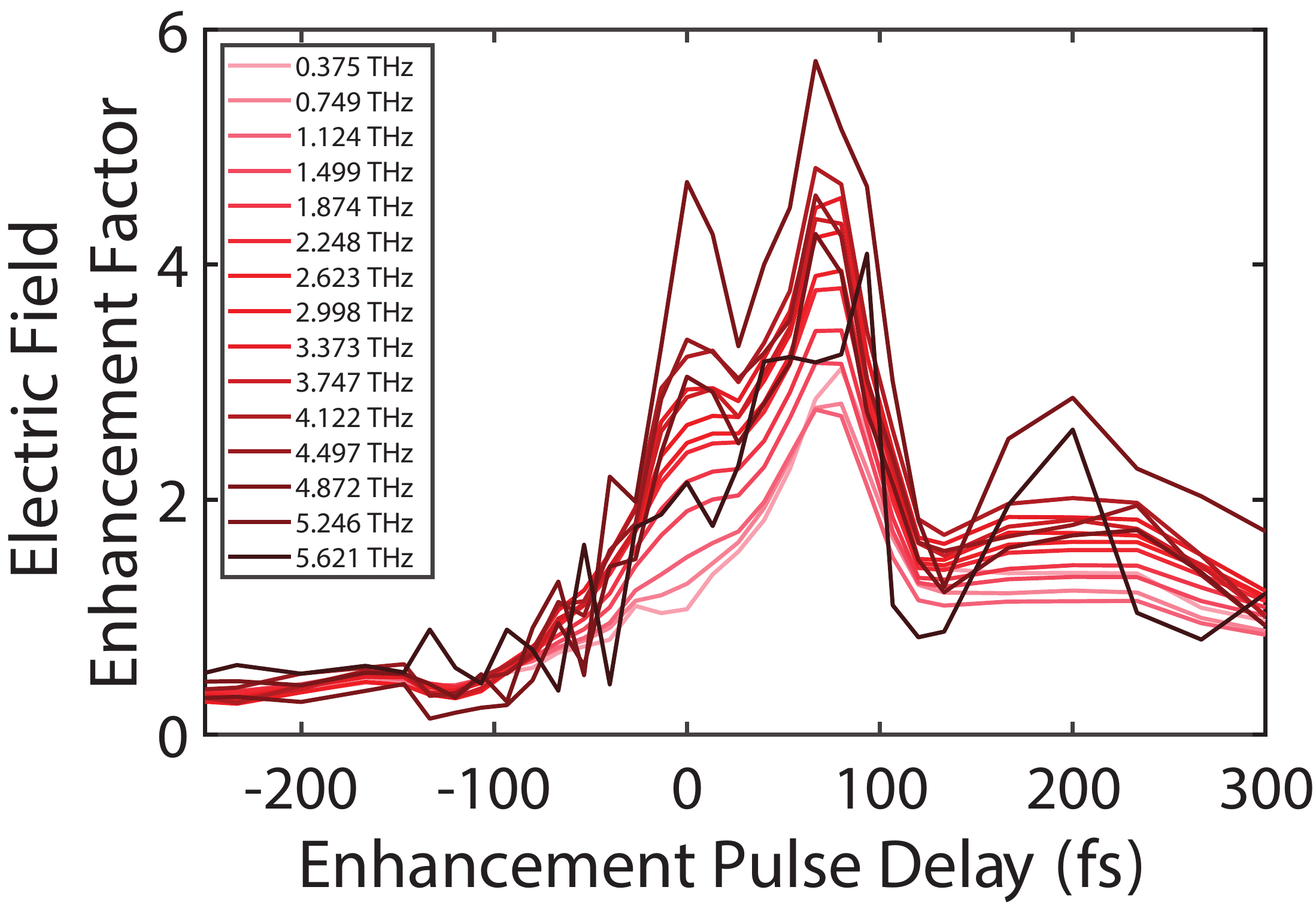}\\
  \caption{\label{figSpectralEnhance} Enhancement Factor traces, similar to those shown in Figs. 1 and 2 in the main text.  Whereas those shown in the main text are calculated based on the integrated THz spectra, here are shown several such traces derived at a single THz frequency.  The general shape and magnitudes of each are in agreement demonstrating that there is no strong dependence of the enhancements discussed in the main text on the frequency of the emission (within our detected bandwidth).}
\end{figure}

Our assumption that the plasma frequency does not affect the electric field enhancement factors presented in the manuscript is further verified by calculating spectrally-resolved, enhancement-factor traces.  This is shown in Figs. \ref{figSpectralEnhance} and \ref{figSpectralEnhance2D} where it can be seen that the general shape and magnitude of the enhancement factor traces are not sensitive functions of the emission frequency.  We extrapolate this to mean that the shape and magnitude of the enhancement factor are also not sensitive functions of the plasma frequency.  Fig. \ref{figSpectralEnhance} shows experimental traces of the Electric Field Enhancement Factor at various THz frequencies.  It is seen that the general shape of the enhancement factor is consistent for all of the frequencies and the magnitudes agree within a factor of 2.  This demonstrates that there is no strong frequency dependence of the enhancement factors.  

\end{document}